\documentclass[12pt]{article}
\usepackage{epsfig,enumerate,verbatim}
\usepackage{graphicx}
\begin{document}

\begin{center}
{\large \bf The large nonlinearity scale limit of an information-theoretically motivated nonlinear Schrodinger equation}
\end{center}
\vspace{0.1in}

\begin{center}

{Le-Huy Nguyen, Hai-Siong Tan and Rajesh R. Parwani\footnote{Email: parwani@nus.edu.sg}}

\vspace{0.3in}

{Department of Physics,\\}
{National University of Singapore,\\}
{Kent Ridge,\\}
{ Singapore.}

\vspace{0.3in}

\end{center}
\vspace{0.1in}
\begin{abstract}
A nonlinear Schrodinger equation, that had been obtained within the context of the maximum uncertainty principle, has the form of a difference-differential equation and exhibits some interesting properties. Here we discuss that equation in the regime where the nonlinearity length scale is large compared to the deBroglie wavelength; just as in the perturbative regime, the equation again displays some universality. 
We also briefly discuss stationary solutions to a naturally induced discretisation of that equation. 
\end{abstract}

\vspace{0.5in}

\section{Introduction}
Physicists study nonlinear Schrodinger equations for two purposes: (i) as effective equations in such fields as condensed matter \cite{sulem} and (ii) to probe a potential deviation from exact linearity in quantum mechanics \cite{nl}. 

The manner in which the nonlinear Schrofinger equations are arrived at in the literature is varied. Here we will focus on the nonlinear equation that was obtained using information theoretic arguments \cite{P1}. Such an approach, often termed the ``maximum entropy method" or more generally ``maximum uncertainty method", is often used in statistical mechanics to deduce the form of the probability distribution. The philosophy is that one should choose the probability distributions with minimum bias while satisfying relevant constraints \cite{J}.

The maximum uncertainty approach had been used to understand the usual linear Schrodinger equation \cite{reg}. Of course in such an approach one must motivate the information, or inverse uncertainty, measure that is used to quantify the entropy (uncertainty) associated with the probability distribution. Just as the Gibbs-Shannon measure is the simplest one that satisfies certain axioms in statistical mechanics, the Fisher measure emerges as the simplest one that satifies axioms appropriate for classical ensemble dynamics, and hence for deducing non-relativistic quantum mechanics as a single parameter extension of classical ensemble dynamics \cite{P2}. 

One could use the approach of \cite{P2} to obtain generalisations of the Fisher measure, involving higher derivatives, that satisfy the same axioms but which involve more parameters. However the natural constraint that the equation be invariant under a scaling of the wavefunction then leads to nonpolynomial terms $\sim \partial p / p $, with $p(x)$ the probability density; and since the nonlinearity is expected to be phenmenologically small, the nonlinear corrections would lead generally to ill-defined singular terms in perturbation theory. 

A different approach to generalising the Fisher measure is to choose the simplest nonsingular measure that contains the Fisher measure in some limit, satisfies the various axioms and which also has some physical motivation. That is how the regularised Kullback-Lieber measure came to be used and the associated nonlinear Schrodinger equation derived \cite{P1}.

For simplicity we consider here only a single quantum particle in one space dimension, see \cite{P1} for the general case. The nonlinear equation is  

\begin{equation}
i \hbar {\partial \psi \over \partial t} = - {{\hbar}^2 \over 2m} {\partial^2 \psi \over \partial x^2} + V(x) \psi + F_1(p) \psi \, , \label{nsch1}
\end{equation}
with
\begin{eqnarray} 
  F_1(p) &\equiv& Q_{1NL} - Q \\
\end{eqnarray}
and $p(x) = \psi^{\star}(x) \psi(x)$; also  
\begin{equation}
{\cal{E}} L^2 = {{\hbar}^2  \over 4 m } \, , \label{uncert}
\end{equation}
and
\begin{eqnarray}
Q &=&  - {{\hbar}^2 \over 2m}  {1 \over \sqrt{p}} {\partial^2 \sqrt{p} \over \partial x^2} \, \; . \label{pot1} 
\end{eqnarray}
Let 
\begin{eqnarray}
p_{\pm}(x) & \equiv & p(x \pm \eta L) \,  ,
\end{eqnarray}
where the dimensionless parameter $\eta$ takes values $0 < \eta \le 1$, then 
\begin{equation}
Q_{1NL}= { {\cal{E}}  \over \eta^4}  \left[ \ln {p \over (1-\eta) p + \eta p_{+} } + 1 - {(1-\eta) p \over (1-\eta) p + \eta p_{+}} - {\eta p_{-} \over (1-\eta) p_{-} + \eta p} \right]  \, . \label{Q2}
\end{equation}

In addition to regulating potential singularities, $\eta$ may also be viewed as parametrising a family of nonlinear theories, with the usual linear quantum mechanics recovered  at $\eta =0$. 

The nonlinear Schrodinger equation (\ref{nsch1}) resembles a differential-difference equation as the evolution of the wavefunction depends, any fixed time, not just on knowledge at the point $x$ but also at neighbouring points a finite distance away, $x \pm \eta L$.
 The nonlinearity is also non-polynomial.  
Nevertheless, the equation shares a number of important properties with the linear Schrodinger equation, such as the conservation of probability and existence of the usual plane wave solutions.

\section{Perturbative Regime}
It was found \cite{P3} that for smooth external potentials $V(x)$, treating the nonlinearity as a perturbation results in energy shifts which depend significantly on whether or not the  unperturbed states have nodes. Those that do, have their energies shifted by a larger amount
\begin{equation}
\delta E \approx {\frac{\hbar ^{2}|L| \pi}{6m}}\ \sqrt{\eta(1 - \eta)}\ \
(1-4 \eta)
\sum_{p=1}^{N}C_{np}^{2}\, + O(L/a)^2 , \label{EPX}
\end{equation}
where $a$ is the characteristic length scale of the linear theory, and $L/a$ is the perturbation parameter.
 The coefficients $C_{np}$ depend on the slope of the unperturbed wavefunctions near the nodes, and hence on the external potential.
 Notice that the dependence of the energy shifts on $\eta$ is universal, resulting in positive  energy shifts for small $\eta$ and negative shifts for larger values. 

One may also try to fix the value of $\eta$ \cite{P4}. If one assumes $\eta$ to be a parameter that ``flows" in the space of theories, then one may see whether there is some value that minimises the energy of the system. Indeed, the energy shifts mentioned above reach a unique global minimum at 
\begin{equation}
\eta_m \approx 0.80 \, .
\end{equation}
At this value of $\eta$ the leading energy shifts are {\it negative} so that the nonlinearity reduces the energy of the original linear system.

\section{Nonperturbative regime}
For $L$ large, $p(x\pm L) \sim p(L) \sim 0$, and so the time-independent nonlinear Schrodinger equation for an eigenstate reduces to 
\begin{equation}
E\phi(x) =V(x)\phi(x) - {{\cal{E}} \over \eta^4} \log(1-\eta) \phi(x) \, \label{shrink},
\end{equation}
where $V(x)$ is a given external potential. 
This equation suggests that $\phi(x)$ can be nonzero only at those points where the potential vanishes. For a potential with a single minimum, this means that the wavefunctions collapse to a delta function and the energy is given in this limit by 

\begin{equation}
E= - {{\cal{E}} \over \eta^4} \log(1-\eta)
\end{equation}
As a function of $\eta$ this energy attains a minimum at the universal, that is $V(x)$ independent, value 
\begin{equation}
\eta \sim 0.9. 
\end{equation}
Notice also that because of (\ref{uncert}), the energy goes to zero as $L \to \infty$.

A variational calculation gives a quick, though approximate, interpolation between the small and large $L$ regimes. 
We did this for the SHO potential using trial states which are SHO stationary states but with the usual deBroglie wavelength $a= \sqrt{{\hbar \over m \omega}}$ replaced  by a variational parameter $b$ \cite{varr}. The dimensionless quantity $c=b/a$ was  varied to minimise the action, which depends on the free parameters $\eta$ and $\epsilon =L/a$. The energy functional was then evaluated at the minimum for each $(\eta, \epsilon)$.  
 
 For small $\epsilon$ the results are in agreement with the perturbative results as expected. For larger $\epsilon$ we found that the wavefunction shrinks and expands suddenly at certain ($\eta$ dependent) values of $\epsilon$ before eventually shrinking ($c\to 0$) to a highly localised state as suggested by (\ref{shrink}).  The oscillatory behaviour of the wavefunction might be an artifact of the variational calculation. Figures (1-3) summarise some of the variational results. (The dependence of the energy on $\eta$ is qualitatively similar to that in the perturbative regime).

 \section{Discretised States}
 Since the nonlinear equation already has the form of a difference-differential equation, it is most natural to consider the fully discretised equation, with $p(x+L)=p_{n+1}$ etc., as another approach to study nonperturbative aspects of the equation \cite{P5}. We remark that if $L$ is related to gravity, as speculated in \cite{P1}, then this discretisation of space suggested by the structure of the equation is consistent with discretisations implied by other approaches to quantum gravity. 

Consider first the the time-independent nonlinear equation for $\eta =1$. We have  
\begin{eqnarray} 
 E 
&=& {\cal{E}} \left[ \ln {p_n \over p_{n+1}}  \ + 1 \ - {p_{n-1} \over p_n} \right] \, . \label{dis}
\end{eqnarray}

If we seek solutions on the full infinite line for which $p_n$ is bounded for all $n$, then by using the iterative nature of (\ref{dis}) one easily shows that that this is only possible for $E>0$, and even then there is a lower bound on $n$ (to keep the probability positive). {\it That is, the discretised space is necessarily truncated}. 
Similar results hold for the general regularised case, $\eta \neq 1$, and also for the $q-$ deformed equation in Ref.\cite{P1}.

\section{Summary and Outlook}
The fact that the nonlinear equation (\ref{nsch1}) was motivated through a philosophically  appealing information theoretic approach suggests that the equation might eventually be of physical relevance, for example as an effective equation in one or both of the domains mentioned in the first paragraph of the introduction. As such, it is of interest to investigate the properties of that equation.

We found that for large $\epsilon =L/a$ the nonlinearity acts as a strongly attractive force, causing a localisation of the wavefunctions. The resultant energy of all eigenstates is minimised at a universal value of $\eta \sim 0.9$, which is larger than the value $\eta \sim 0.8$ found (for states with nodes) in the perturbative regime of $\epsilon$.
A natural question is whether for any $L$, there is a $V(x)$-independent $\eta$ value for which the energy of eigenstates is minimised.

There are several other properties of the nonlinear equation that have yet to be studied. Of particular interest are its time-dependent states: dynamics of wavepackets and possible solitary waves. We invite interested readers to the adventure.

{\bf Acknowledgement}
R.P. thanks the organisers for their hospitalty and for staging a stimulating event.

\newpage

\section*{Figure Captions}

\begin{itemize}
\item  Figure 1 : The variationally computed energy versus $\epsilon$. Plots are similar for the ground state and the example of an excited state,$n=5$, studied here. Generally, for fixed $\epsilon$, the energy decreases as $\eta$ increases. 

\item  Figure 2 : The normalised width $c$ of the variational ground state versus $\epsilon$. Looking at the intercept on the x-axis from the right, the curves correspond to the following values of $\eta: 0.1, 0.2, 0.999999, 0.5, 0.999, 0.9. $

\item  Figure 3 : The normalised width $c$ of the variational excited ($n=5$) state versus $\epsilon$.

\end{itemize}

\newpage

\section*{Figures}
\begin{figure}[h]
  \begin{center}
   \epsfig{file=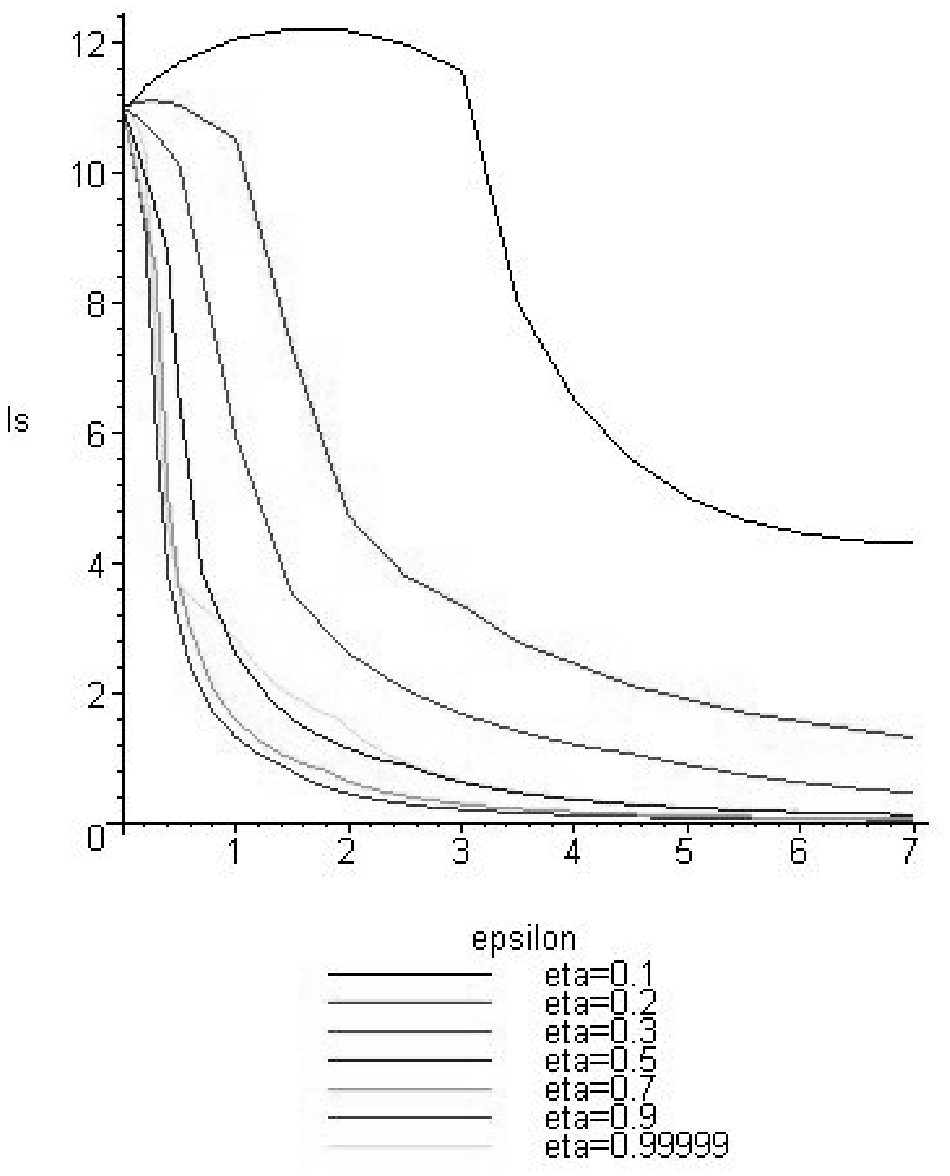, width=9cm}
    \caption{.}
    \label{energy}
  \end{center}
\end{figure}

\begin{figure}
  \begin{center}
   \epsfig{file=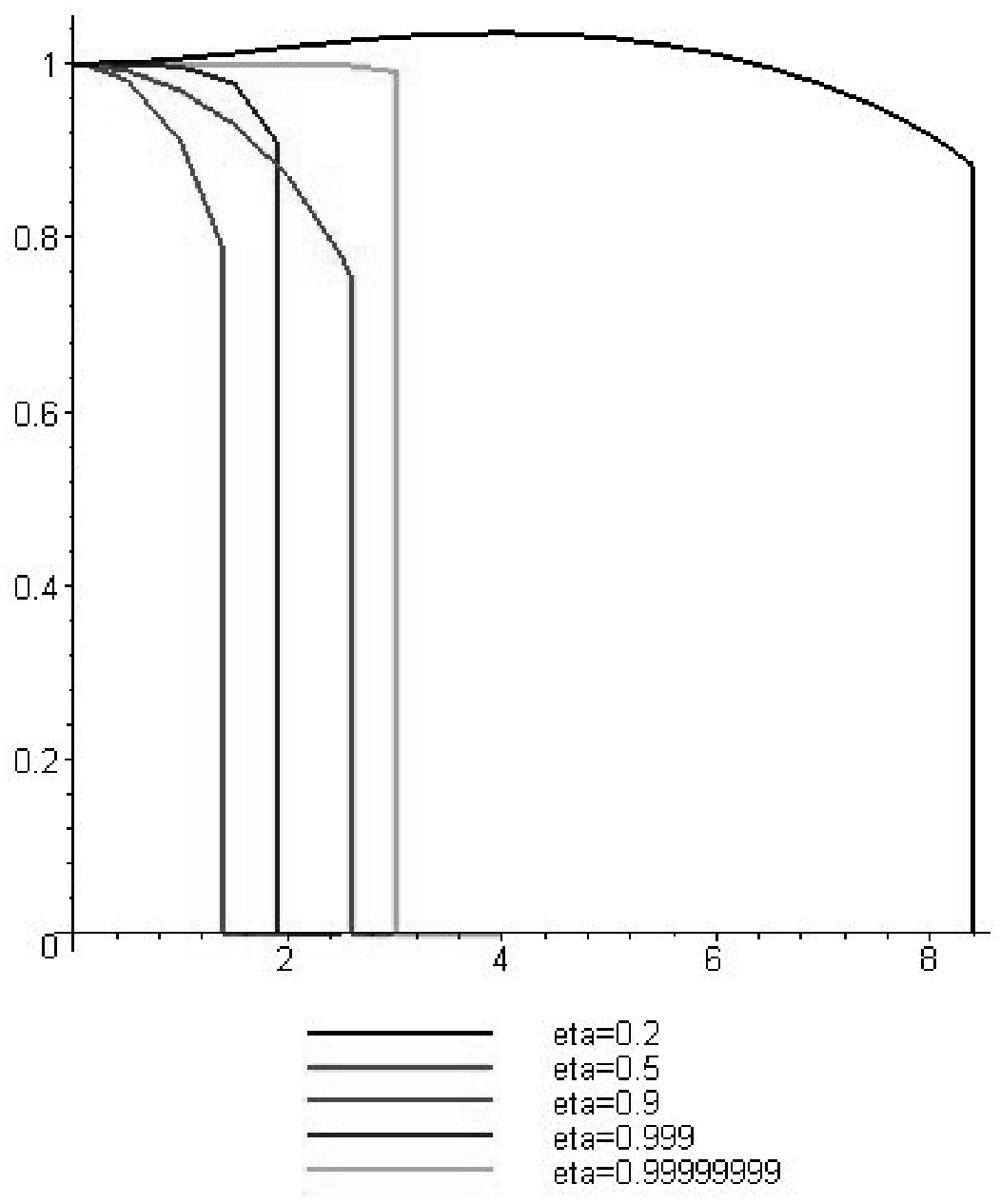, width=9cm}
    \caption{.}
    \label{energy}
  \end{center}
\end{figure}

\begin{figure}
  \begin{center}
   \epsfig{file=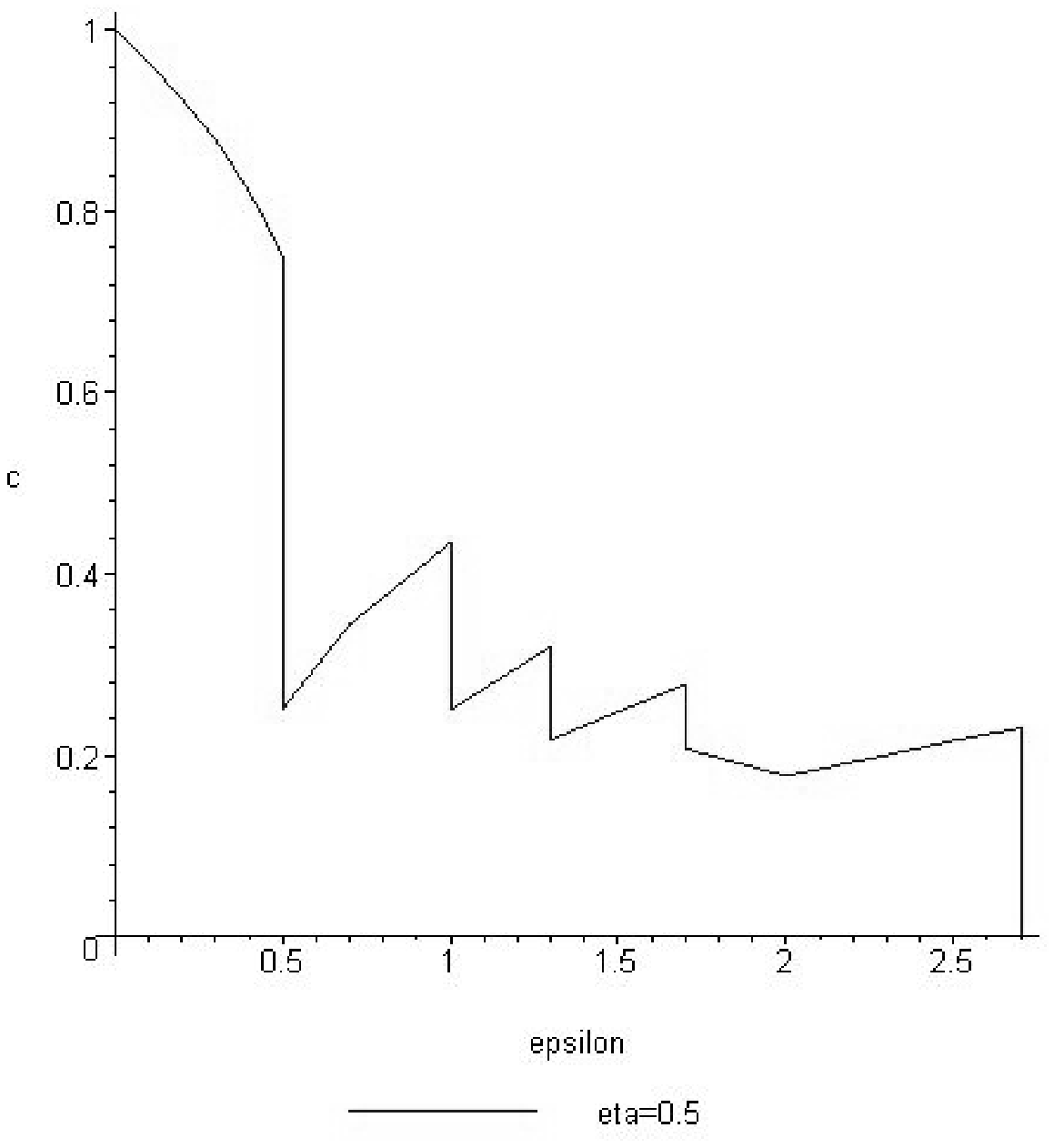, width=9cm}
    \caption{.}
    \label{energy}
  \end{center}
\end{figure}


\begin{thebibliography}{99}

\bibitem{sulem}  C. Sulem and P. Sulem \textit{The nonlinear Schrodinger
equation:self-focusing and wave collapse} (Springer, 1999).

\bibitem{nl} For some reviews, see G. Svetlichny, arXiv:quant-ph/0410036;
R. Carroll,
arXiv:quant-ph/0401082 and references therein.

\bibitem{P1} R. Parwani,  Ann. Phys. \textbf{315}, 419 (2005). 

\bibitem{J}  E.T. Jaynes, Phys. Rev. \textbf{106}, 620 (1957); \textbf{108}%
, 171 (1957); \\
\textit{Probability Theory, The Logic of Science} (Cambridge University
Press, 2004).

 \bibitem{reg} M. Reginatto, Phys. Rev. A58, 1775 (1998); Erratum {\it ibid}. A60 1730 (1999); \\
B. R. Frieden, J. Mod. Opt 35,1297 (1988); Am. J. Phys. 57 (1989) 1004.\\
 
\bibitem{P2} R. Parwani, J. Phys. A:Math. Gen. \textbf{38}, 6231 (2005);\\
R. Parwani,  Int. J. Theor. Phys. \textbf{45}, 1901 (2006).

\bibitem{P3} R. Parwani and G. Tabia, J. Phys. A: Math. Theor. 40 5621-5635 (2007).  

\bibitem{P4} R. Parwani, Theoretical and Mathematical Physics, 152(1): 1012–1016 (2007).

\bibitem{varr} R. Hasson and D. Richards, arXiv:quant-ph/0012068v1.


\bibitem{P5} R. Parwani and H. S. Tan, Phys.Lett.A363:197-201 (2007).\\






\end{thebibliography}
\end{document}